\documentstyle[psfig]{l-aa}
\begin{document}
\def\tiret{\multicolumn{1}{c}{\mbox{---}}}
\newcommand{\etal}{et al.}
\newcommand{\aj}{AJ}
\newcommand{\apj}{ApJ}
\newcommand{\apjs}{ApJ Supplt}
\newcommand{\apjl}{ApJ Letters}
\newcommand{\aap}{A\&A}
\newcommand{\aaps}{A\&AS}
\newcommand{\gca}{Geochim. Cosmochim. Acta}
\newcommand{\pasp}{PASP}
\newcommand{\qjras}{DJRAS}
\newcommand{\solphys}{Sol. Phys.}


\thesaurus{
02.08.1; 
08.09.3, 
08.18.1. 
}

\title{Mixing processes during the evolution of red giants with 
moderate metal deficiencies : the role of molecular-weight 
barriers\thanks{Based partially on observations obtained at the 
Dominion Astrophysical Observatory, National Research Council, Canada}
}

\author{Corinne Charbonnel\inst{1,2}, Jeffery A. Brown\inst{3,4}, 
George Wallerstein \inst{3,5}}

\offprints{Corinne Charbonnel}

\institute{Laboratoire d'Astrophysique de Toulouse, CNRS UMR 5572,
Toulouse, France
\and 
Space Telescope Science Institute, 3700 San Martin Drive, Baltimore,
MD 21218, USA
\and
Dept. of Astronomy, University of Washington, Box 351580, Seattle, WA
98195-1580, USA
\and 
Program in Astronomy, Washington State
University, Pullman, WA 99164-3113, USA
\and Guest observer, Dominion Astrophysical Observatory, Herzberg 
Institute of Astrophysics, National Research Council, Canada
\\
Corinne.Charbonnel@obs-mip.fr, jbrown@delta.math.wsu.edu,
wall@astro.washington.edu}

\maketitle
\markboth{Charbonnel, Wallerstein and Brown}{Mixing Processes on RGB}

\date{Received / Accepted }

\begin{abstract}

We have assembled accurate abundance data for Li, C, and N as well 
as the $\rm ^{12}C/^{13}C$ ratio for five field giants 
with [Fe/H] $\simeq - 0.6$ including Arcturus 
and two stars in the globular cluster 47 Tuc.
Using their very precise $\rm M_{bol}$ values obtained from HIPPARCOS
parallaxes, we can place them into an evolutionary sequence.
The sequence shows that
the $\rm ^{12}C/^{13}C$ ratios drops from $\sim 20$ to near 7 between
$\rm M_{bol}$ = +1 and +0.5, while Li disappears.
At the same time the $\rm ^{12}C/^{14}N$ ratio diminishes by 0.2 to 0.4 
dex.  
The two stars in 47 Tuc with Mbol near -2.0 show even lower 
$\rm ^{12}C/^{14}N$ ratios by 0.4 dex indicating further mixing as they 
evolved to the top of the red giant branch.

These observations confirm the existence of an extra-mixing process that
becomes efficient on the red giant branch only when the low-mass stars 
reach the so-called luminosity function bump.
We use the values of the carbon isotopic ratio observed in our sample
to get constraints on the $\mu$-barriers that may shield the central 
regions of a star from extra-mixing.
We show that the same value of the critical gradient of molecular weight
leads to $\rm ^{12}C/^{13}C$ ratios observed at different metallicities.
This ``observational critical $\mu$-gradient''
is in very good agreement with the
one which is expected to stabilize meridional circulation.
This result provides strong clues on the nature of the extra-mixing which
occurs on the RGB, and indicates that it is related to rotation.


\end{abstract}

\section{Introduction}

For some time now it has been clear that evolved low mass stars
exhibit chemical anomalies which are not predicted
by standard stellar evolution theory.
By ``standard" we refer to the hypotheses that stellar convective regions
are instantly mixed and that no transport of chemicals occurs in the
radiative regions.
In addition, standard models are those without rotation at any depth.
With these assumptions, changes in the surface abundances
prior to the asymptotic giant branch (AGB) stage
are only expected to be due to convective dilution
during the first dredge-up phase.
In low mass red giant branch (RGB) stars, the convective envelope reaches
only regions where $^{12}$C was processed to $^{13}$C and $^{14}$N.
Consequently, the carbon isotopic ratio declines (from 90 to about
20-30), the carbon abundance drops (by about 30\%) and nitrogen
increases (by about 80\%), but oxygen and all heavier element abundances
remain unchanged.
According to the standard scenario, the surface abundances then
stay unaltered as the convective envelope slowly withdraws
outward in mass
during the end of the RGB evolution.

However, observational data
reveal a different reality.
The first discrepancy came among the first stellar $\rm ^{12}C/^{13}C$
ratios published (Day et al. 1973), when Arcturus was found to have
$\rm ^{12}C/^{13}C = 7.2 \pm 1.5$.
The discrepancy  was aggravated by
the analysis by Lambert \& Ries (1981) of C, N, and O abundances
in a sample of red giants, including Arcturus.
In fact, in most of the metal-deficient field and globular cluster evolved
stars, the observed conversion of $^{12}$C to $^{13}$C and $^{14}$N
greatly exceeds the levels expected from standard stellar models;
the $\rm ^{12}C/^{13}C$ ratio even reaches the near-equilibrium value
in many Pop II RGB stars
(Sneden, Pilachowski \& Vandenberg 1986; Smith \& Suntzeff 1989;
Brown \& Wallerstein 1989, 1992; Gilroy \& Brown 1991, henceforth GB91;
Brown, \etal 1990, henceforth BWO;
Bell et al. 1990; Suntzeff \& Smith 1991;
Shetrone et al. 1993, henceforth SSP93; Briley et al. 1994, 1997).
This problem actually also occurs, but to a somewhat lower extent,
in evolved stars belonging to open clusters with turnoff masses
lower than 2M$_{\odot}$ (Gilroy 1989; GB91).

In addition,
Population II evolved stars present other chemical anomalies.
In halo giants, the lithium abundance continues to decrease
after the completion of the first dredge-up (Pilachowski et al. 1993).
A continuous decline in carbon abundance with increasing stellar
luminosity along the RGB is observed in globular clusters such as
M92 (Carbon  1982; Langer \etal 1986),
M3 and M13 (Suntzeff 1981),
M15 (Trefzger \etal 1983),
NGC 6397 (Bell \etal 1979, Briley \etal 1990),
NGC 6752 and M4 (Suntzeff \& Smith 1991).
In some globular clusters
(M92, Pilachowski 1988;
M15, Sneden \etal 1991;
M13, Brown \etal 1991, Kraft \etal 1992;
$\omega$ Cen, Paltoglou \& Norris 1989),
giants exhibit evidence in their atmospheres for O$\rightarrow$N
processed material.
In addition to the O versus N anticorrelation, the existence of Na
and Al versus N correlations and Na and Al versus O anticorrelations in a large
number of globular cluster red giants has been clearly confirmed
(Drake \etal 1992, Kraft \etal 1992, 1993, Norris \& Da Costa 1995,
Kraft 1994, Kraft \etal 1997, Shetrone 1996, Zucker \etal 1996).

These observations suggest that low mass stars undergo a non-standard mixing
process which adds to the first dredge-up and modifies the surface abundances.
The first and only evidence on the evolutionary state at which this
non-standard mixing actually becomes effective
came from observations of
the C/N and $\rm ^{12}C/^{13}C$ ratios
(respectively by Brown 1987 and by GB91)
in evolved stars of M67.
In this old open cluster, which has a turnoff mass of 1.2 M$_\odot$
(VandenBerg 1985, Meynet \etal 1993)
and a value of [Fe/H]=0.0$\pm$0.1,
the observational and theoretical first dredge-up appear to be in complete
agreement (Charbonnel 1994, henceforth C94).
Indeed, the $\rm ^{12}C/^{13}C$ ratio changes at the base of the giant
branch from a value $>$40 to 21-25 between $\rm M_v \approx +3.5$ and $\rm M_v
\approx +2.5$; C/N changes by about a factor of 4 in this same 
interval.
The onset of the dredge-up occurs at a luminosity in accord with standard
theory,
and the observed post-dilution ratios are in very good agreement
with the standard predictions at this point.
However, in M67 an additional mixing event begins
at about $\rm M_v \approx +0.5$ where $\rm ^{12}C/^{13}C$
drops from 21-25 to 11-15.
This luminosity actually corresponds to the so-called ``bump" in the
RGB luminosity function (Fusi-Pecci \etal 1990), i.e.,
to the evolutionary point where the hydrogen burning shell crosses the
chemical discontinuity created by the convective envelope at its maximum
extent.
As pointed out by C94, this observational fact strongly suggests that prior
to this evolutionary point the mean molecular weight gradient created during
the first dredge-up acts as a barrier to any mixing below the convective
envelope.
After this point, however, the gradient of molecular weight above the
hydrogen-burning shell is much lower, and extra-mixing is free to act.

Since the observations in M67, there have been no new sets of abundance data
in any cluster, either globular or galactic,  to establish the
evolutionary point at which the extra-mixing begins to occur.
With 4-meter class telescopes it is almost impossible to observe
red giants in globular clusters at the luminosities at which
$\rm ^{12}C/^{13}C$ is expected to drop from near 20 to less than 10 because
the molecular bands become weaker with increasing temperature and gravity.
Furthermore, the attainable signal-to-noise of the spectra diminishes with
decreasing stellar luminosity.
It is crucial, however, for our understanding of the nature of the
process to compare conditions in stars of different populations and 
metallicities.

To this end, we address here the problem of the onset of the 
extra-mixing, and of its possible metallicity dependance, 
by placing in an evolutionary sequence five field giants with moderate 
metal deficiencies and two giants of the globular cluster 47 Tuc for 
which we have
assembled accurate abundance data for Li, C and N as well as for the
$\rm ^{12}C/^{13}C$ ratio.
Our observational data are presented in \S 2.
We enlarge our sample by including in our discussion the old disk giants
with $\rm ^{12}C/^{13}C$ ratios derived by SSP93.
Very precise absolute magnitudes were recomputed from HIPPARCOS parallaxes
for all the stars we consider.
As discussed in \S 3, the sequence allows us to confirm that the
discrepancy between observations of mixing-sensitive species and
standard theory really appears at the luminosity-function bump (LFB) 
on the RGB, but not at lower luminosities. 
Since our sample is moderately metal-deficient, we can
study the metallicity-dependence of the extra-mixing by comparing
our results with the observations in M67.
In \S 4, we investigate the nature of the factor that stabilizes the star
against extra-mixing before it reaches the luminosity-function bump,
and discuss the role of molecular-weight ($\mu$) barriers.
From simple considerations, we derive from our data the ``observational"
critical $\mu$-gradient, ($\nabla \ln \mu)_{c,obs}$, that shields the 
central regions of a star from extra-mixing. If we assume that
extra-mixing is free to act down to the layers defined by the same value of
($\nabla \ln \mu)_{c,obs}$ whatever the stellar mass and metallicity, then
$\rm ^{12}C/^{13}C$ ratios can be explained both for Pop I and II giants.
Finally, we discuss the implications for the nature of the extra-mixing
mechanism itself on the RGB, and bring clues that it may be related to
rotation-induced processes.

\section{Observational data for metal poor stars}

\begin{table*}
\caption{Observed Stars $^a$}
\begin{tabular}{cccccccc}
\hline \\[1mm]
\multicolumn{1}{c}{HD}&
\multicolumn{1}{c}{Star}&
\multicolumn{1}{c}{V - K}&
\multicolumn{1}{c}{T$_{eff}$}&
\multicolumn{1}{c}{M$_V$}&
\multicolumn{1}{c}{M$_{bol}$}&
\multicolumn{1}{c}{log g}&
\multicolumn{1}{c}{[Fe/H]} \\[1mm]
\hline \\[1mm]
37160 & $\phi^2$ Ori &2.41& 4720 & $+1.33\pm$ 0.07 & $+$0.89 & 2.45 &
-0.70\\
3546  & $\varepsilon$ And&2.13& 4925 & $+0.77 \pm$ 0.09 & $+$0.43 & 2.34 
&$-$0.66\\
89484&$\gamma$Leo A&\tiret &4440&$-0.32\pm$0.07 &$-$0.9&1.62& $-$0.46\\
89485&$\gamma$Leo B&\tiret &4900&$+0.87\pm$0.07 &$+$0.5&2.36&$-$0.42\\
124897& Arcturus & 3.00 & 4330 & $-0.30\pm$0.02 &$-$0.98& 1.55 & $-$0.62\\
\tiret & 47 Tuc 3501 & 3.45 & 3975 & $-$1.1 & $-$1.9 & 0.9 & $-$0.65 \\
\tiret & 47 Tuc 4418 & 3.50 & 3975 & $-$1.1 & $-$1.9 & 0.9 & $-$0.65 \\
[1mm] \hline
\end{tabular}

$^a$ The stars were selected so as to be as close to the evolutionary 
tracks delineated by the color-magnitude diagram of 47 Tuc as their
parallaxes (ground based when we selected the stars) and [Fe/H] values
from the literature permitted
\end{table*}

\begin{table*}
\caption{Comparison of our abundances (upper analysis) with those 
of Cottrell and Sneden (1986) and Shetrone et al.(1993) (lower
analysis) of $\varepsilon$ And and $\phi^2$Ori}
\begin{tabular}{ccccccccc}
\hline \\[1mm]
\multicolumn{1}{c}{Star} &
\multicolumn{1}{c}{$\rm T_{eff}$} &
\multicolumn{1}{c}{log g} &
\multicolumn{1}{c}{[Fe/H]} &
\multicolumn{1}{c}{[C/Fe]} &
\multicolumn{1}{c}{[N/Fe]} &
\multicolumn{1}{c}{[O/Fe]} &
\multicolumn{1}{c}{$\rm ^{12}C/^{13}C$} &
\multicolumn{1}{c}{$\rm \log \epsilon(Li)$} \\[1mm]
\hline \\[1mm]
$\varepsilon$ And &4925 &2.55 &$-$0.45 &$-$0.29 &+0.24 &$-$0.06 &7 &$<$ 
$-$0.2\\
 &4750$^{a,b}$ &2.0$^{a,b}$ &$-$0.7$^{a,b}$ &$-$0.15$^b$
&$-$0.10$^b$ &+0.2$^b$ &7.5$^a$ &0.0$^b$ \\
$\phi^2$ Ori &4720 &2.85 &$-$0.43 &$-$0.03 &+0.08 &+0.15 &$>$20 &+0.3\\
 &4600 &2.4 &$-$0.7 &0.00 &$-$0.20 &+0.15 &$>$40 & \\
\hline
\end{tabular}

$^a$ ~~From Shetrone \etal\ (1993)

$^b$ ~~From Cottrell and Sneden (1986)
\end{table*}

Some of the data that we employ here have already been published.
We concentrate on stars of modest metal deficiency.
Five bright metal-deficient field stars are
$\phi^2$ Ori, $\varepsilon$ And, $\gamma$ Leo B, $\gamma$ Leo A
and Arcturus.
Their $\rm M_{bol}$ values range from near +0.9 to $-$1.0 so they form an
evolutionary sequence that starts from fainter absolute magnitude
than can be reached with high spectral resolution in globular clusters.
As a continuation of our ``evolutionary sequence" we include two stars from the
globular cluster 47 Tucanae whose properties indicate that they are first giant
branch (i.e. not AGB) stars.
The basic properties of our sample stars are shown in Table 1.

\begin{table*}
\caption{Abundance Results}
\begin{tabular}{ccccccccc}
\hline \\[1mm]
\multicolumn{1}{c}{Star}&
\multicolumn{1}{c}{$\rm ^{12}C$} &
\multicolumn{1}{c}{$\rm ^{13}C$} &
\multicolumn{1}{c}{ N} &
\multicolumn{1}{c}{ O} &
\multicolumn{1}{c}{$\rm (C + N)^a $} &
\multicolumn{1}{c}{$\rm (C + N + O)^b $} &
\multicolumn{1}{c}{$\rm ^{12}C/^{13}C $} &
\multicolumn{1}{c}{log $\epsilon$(Li) } \\[1mm]
\hline \\[1mm]
$\phi^2$ Ori& 7.9 & $<$6.6 & 7.8 & 8.55 & 8.17 & 8.7 & $>$20$^c$ & 
0.3 \\
$\varepsilon$ And& 7.65 & 6.8 & 7.95 & 8.3 & 8.15 & 8.53 & 7$^d$ & 
$< -0.20$$^e$\\
$\gamma$ Leo B& 7.77 & 6.8 & 7.87 & 8.45 & 8.14 & 8.62 & 9.5 & $\le 
+0.3$$^f$ \\
$\gamma$ Leo A& 7.8 & 7.0 & 7.9 & 8.45 & 8.18 & 8.64 & 6.5 & $\le 
0.0$$^f$ \\
Arcturus & 7.9 & 7.0 & 7.7 & 8.55 & 8.15 & 8.7 & 7 & $<-1.5$$^g$ \\
47 Tuc 3501 & 7.7 & 6.85 & 8.25 & 8.45 & 8.37 & 8.7 & 8 & 
~~~$^h$ \\
47 Tuc 4418 & 7.6 & 6.75 & 8.25 & 8.5 & 8.35 & 8.7 & 7 & 
~~~$^h$ \\
$\odot$$^i$  & 8.6 & 6.6 & 8.0 & 8.9 & 8.7 & 9.1 & 90 & 1.15\\
\hline
\end{tabular}

$^a$ ~~This is $\rm {^{12}C} + {^{13}C} + N$

$^b$ ~~This is $\rm ^{12}C + {^{13}C} + N + O$

$^c$ ~~Shetrone \etal 1993 find $\rm ^{12}C/^{13}C > 40$

$^d$ ~~Shetrone \etal 1993 find $\rm ^{12}C/^{13}C = 7.5$

$^e$ ~~Shetrone \etal 1993 find Li = 0.0

$^f$ ~~Helfer \& Wallerstein 1968

$^g$ ~~Lambert \etal 1980

$^h$ ~~not present, no upper limit computed

$^i$ ~~photospheric value, ~~Anders \& Grevesse 1989, Grevesse \etal 1991
\end{table*}

For the field stars we derive the gravity values
assuming a mass of 0.8 $\rm M_{\odot}$, the HIPPARCOS parallaxes, 
$\rm T_{eff}$'s derived from the $\rm V - K$ color-$\rm T_{eff}$ relation
(Ridgway \etal 1980, DiBenedetto 1993, Dyck \etal 1996),
and the bolometric corrections of Buser and Kurucz (1992).
This method is probably superior to spectrocopic determination
of gravities using ionization balance; a test using the HIPPARCOS
parallaxes indicates the spectroscopic gravities
often are too low by a factor of two (Nissen \& H{\o}g 1997).
The visual binary stars in $\gamma$ Leo do not have separate colors;
instead, we adjust the spectroscopic $\rm T_{eff}$'s of Lambert \& Ries (1981)
downward by 210K, the mean offset between their temperature scale and
the $\rm V - K$ scale.
We note that the 47 Tuc stars are probably not on the same absolute 
magnitude scale as the HIPPARCOS results, since they depend upon a
cluster distance modulus which probably needs revision in light of
the HIPPARCOS results for the subdwarfs;
see, {\it e.g.,} Reid (1997).
For our purposes, the precise value of the absolute magnitudes of the two 
globular cluster stars is not important:
it matters here only that they are $\sim 1$ mag more luminous than 
our most luminous field stars.
\footnote{Both Arcturus and Gamma Leo A lie above the horizontal
branch and hence could be post-He-flash stars. 
However their colors place them on the first ascent giant branch. 
A comparison of the colors we have quoted shows that Arcturus falls 
exactly on the first ascent giant branch of 47 Tuc (Hesser et al. 1987). 
For Gamma Leo the measured B-V color pertains to both stars and must
be deconvolved to obtain the color of star A. To derive the observed B-V
color of 1.15 mag, we combined the color of Arcturus, B-V=1.23 with that of
the G8III secondary, 0.95 mag, using the observed 1.2 visual magnitude 
difference, to derive the observed composite color of 1.15 mag. 
This confirms that the color of Gamma Leo A is about the same as that of 
Arcturus and hence that Gamma Leo A on the first ascent giant branch.}

The abundances we use have been rederived with a number of improvements.
For Arcturus and the $\gamma$ Leo pair we have combined the equivalent widths
of Lambert \& Ries (1981) and Ries (1981) with 
the empirical photometric $\rm T_{eff}$ scale
and the
most recent and almost certainly most accurate value of the CN dissociation
potential $\rm D^o_o$ = 7.65 eV (Bauschlicher \etal 1988).
For 47 Tuc we have used the data of BWO with two changes.
The analysis by Hesser \etal (1987)
of the color-magnitude diagram of 47 Tuc indicate an iron abundance
for 47 Tuc higher by about 0.2 dex as compared with the BWO value.
Hence, we have accepted
[Fe/H] = $-$0.65 rather than $-$0.85 for the 47 Tuc stars
and have recalculated the CNO abundances
using models with the new [Fe/H] values.  For the $\rm ^{12}C/^{13}C$ ratio in
47 Tuc star 4418, we have adopted the second solution in
Table 2 of
Bell \etal (1990;
multiple solutions for $\rm ^{12}C/^{13}C$ and carbon abundance
are given for different values of the stellar microturbulence parameter)
because the corresponding derived carbon abundance is closer to
that of BWO (in which the microturbulence is derived as part of the
analysis of the high resolution spectra in the CH region).

To extend our evolutionary sequence to lower luminosity on the giant branch we
have obtained new high resolution spectra of two field stars $\phi^2$ Ori
and $\varepsilon$ And, whose metallicities are similar to those of the other
stars.  The new spectra were obtained at the Dominion Astrophysical
Observatory with the 1.2-meter telescope, coude spectrograph, long camera and
Reticon detector.  This combination yields a resolving power of 35,000, and
the exposures were timed for a signal-to-noise of about 200.  Our spectra
cover the C$_2$ Swan band at 5635 \AA, the [O I] line at 6300 \AA, the Li line
at 6700 \AA~and the 2-0 red system CN band at 8000 \AA,
thereby providing abundances of
Li, C, N, O as well as iron and the ratio $\rm ^{12}C/^{13}C$. These stars
were analyzed using the same models and gf-values as were used in the other
analyses except that the C$_2$ Swan bands replaced the violet CH bands as the
source of the carbon abundance.  To analyze the C$_2$ Swan band at 5635 \AA~it
was necessary to synthesize the spectrum over an interval from 5626-5640 \AA.
The spectroscopic data for the C$_2$ Swan bands is that reviewed by
Grevesse \etal (1991) in their re-evaluation of the solar C abundance.

Both $\varepsilon$ And and $\phi^2$ Ori have been analyzed by others, most
recently with modern model atmospheres and digital spectra by Cottrell \&
Sneden (1986) and subsequently by SSP93.  These
two stars provide an overlap with their large sample of stars of generally
lower luminosity.  We compare their data with ours in Table 2.

Most of the differences between our abundances and those of the Texas group can
be understood as due to differences in effective temperatures and gravities.
Their lower effective temperatures are responsible for their lower iron
abundances and nitrogen abundances because of lower opacities and the
dissociation equilibrium of the CN molecule.  The low masses derived by
SSP93 (Table 6) and of Cottrell \& Sneden (1986, Table 6)
demonstrate
that their gravities are too low by roughly a factor 3 which is the difference
between their and our gravity values for the same stars.

Our final adopted abundances are shown in Table 3.

It is important to understand the uncertainties in the individual abundances
in Table 3.  The iron abundances are almost independent of the CNO abundances
but not entirely so.  Since many lines of Fe~I and several lines of Fe~II are
easily observed the measured line strengths are not an important source of
error.  The uncertainty is surely dominated by the uncertainty in the
atmospheric model, especially at small optical depth where mechanical energy
input and backwarming, such as electron conduction from the
chromosphere, can effect the temperature near the boundary,
especially the  temperature minimum (Kurucz 1996).  
Hence, we are dealing with an uncertain function, T($\tau$), not
just an uncertain parameter.  This problem was evaluated carefully by Leep
\etal (1987) in connection with the metallicity of M71 and the
derivation of solar f-values using either the Holweger-M\"uller (1974) model
or the Bell \etal (1976) model.  The conclusion of that paper was that no
[Fe/H] value 
for a red giant
can be more accurate than $\pm$0.2 and we accept that here.  The
influence of CNO abundances on the [Fe/H] value comes in through the opacity
of CO and CN which effects the atmospheric models.
The infrared CO bands are important for the cooling and structure
of the outermost layers of an atmosphere of a late-type star,
but this effect is less important for stars as warm as those in this study.
The CN lines can contribute substantially
to the atmosphere structure of red giants because the CN red system blankets
a rather large wavelength range near the flux peak,
but CN diminishes in importance quickly as the metallicity of a star is
decreased because the CN column density falls approximately
as metallicity squared while the continuous opacity falls approximately
as metallicity.

The uncertainty in the oxygen abundance is mostly determined by the actual
measurement of the equivalent width of the only useful oxygen line,
$\lambda$6300, in the globular cluster stars, and the second line at
$\lambda$6363 which is barely detectable in the bright field stars.  Since the
oxygen line is formed over a wide range of optical depth the derived abundance
is not very sensitive to the boundary temperature. 
The low carbon abundances in
these stars means that the uncertainty in the C abundance has little effect on
the solution for the
O abundance despite the formation of CO in the atmosphere. 
For oxygen the uncertainties in O/Fe are surely $\pm$0.1 dex or perhaps a 
little lower for Arcturus, for which the measurements are by far the most 
accurate.  
Hence, the somewhat higher O/Fe value for Arcturus, as compared with the 
other stars in Table 3, is probably real.

The formation of CO depletes carbon and an error of 0.10 dex in oxygen will
cause an error of 0.05 dex in the carbon abundance.  Fortunately the CH and
C$_2$ lines are not formed at a small optical depth (as they are in the
sun) but
rather below the levels at which carbon is depleted by CO formation, so the
uncertainty in the boundary temperature is not important.  Considering both the
uncertainty in the oxygen abundance, the actual measurements of CH, and the
presence of some saturation in the CH lines, an uncertainty of 0.15 dex is
possible for the $^{12}$C abundances.

The $\rm ^{12}C/^{13}C$ ratio is determined largely by the clump of $^{13}$CN
lines near 8005 \AA.  This was illustrated in Figure 3 of BWO:
for that star, M13 I-48, a ratio of 6 was derived
and the figure shows that a range from 5 to 8 is possible.  Scaling that
evaluation to Table 2 of this paper indicates that an observed ratio of 7
for three stars could lie between about 5.5 and 9; the pure equilibrium ratio
of 3.5-4.0 is pretty well excluded as are values of 10 or above.

The nitrogen abundance, derived from CN lines, is less dependent on line
measurement uncertainties because many ($\sim$15-20) lines are available
in our spectra.
However, it does depend on the abundances of C and O through the dissociation
equilibrium of CO, CH, and CN, as well as the formation of N$_2$.  Hence the
uncertainties in the C and O abundances come into play as well as the
uncertainty in the CN dissociation potential.  For CN the difference
between the
theoretical value, 7.65 eV (Bauschlicher \etal 1988) and the
recent experimental value of 7.77 $\pm$ .05 eV (Costes \etal 1990)
is disturbing.  We have used the 7.65 eV value.  If the 7.77 eV value is
correct our N abundances must be lowered by 0.2 dex, which lowers our C+N
values by a somewhat smaller amount. 
The reduction in C+N+O introduced by such a
reduction in N is .05 dex for Arcturus and the $\gamma$ Leo pair, and .08 dex
for the 47 Tuc stars.

An error in either the dissociation potential of CN
or the {\it f}-value of the CN red system 2-0 band
would effect all of our nitrogen abundances equally. 
Both of these quantities are not as well determined
as might be hoped: the dissociation energy of CN seems
to be uncertain by about 0.2 eV, and different sources
for the f-values disagree by as much as 30%
(see the discussion in Bauschlicher \etal 1988).
The values used here for these quantities, 7.65 eV for
$D^0_0$ and $8.4 \times 10^{-4}$ for $f_{20}$,
are those used in (for $D^0_0$) or derived from ($f_{20}$)
the analysis of CN red system lines in the solar spectrum by
Sneden \& Lambert (1982), so by construction they
yield the solar nitrogen abundance.
A second source of uncertainty in the N abundance is
introduced by the uncertainty in the C abundance.  For measured CN lines any
error in the C abundance introduces an error of comparable size
and opposite sign in the N
abundance.  Since an error of as much as 0.15 dex in C is possible,
an error of the same size with reversed sign is possible for N.
 
The $\rm ^{12}C/^{13}C$ ratio also has some observational uncertainty,
almost exclusively due to the difference in strength between the 
$\rm {^12}CN$ and
$\rm {^13}CN$ lines,
so that the microturbulent velocity enters into the isotope ratio
solution.  This uncertainty is aggravated as the $\rm {^12}CN$ lines get
stronger and as the isotope ratio increases.
For the stars in this sample, the $\rm {^12}CN$ lines are quite weak:
for $\gamma$ Leo A, which has the strongest CN lines in the sample,
the reduced equivalent widths are in the range
$-5.5 < \log(W / \lambda) < -5.1$,
indicating that saturation (and hence microturbulence-related errors
in the isotope ratio) are minimal.
Our results are summarized in Figures 1 and 3, and discussed below.

\begin{figure}
\centerline{
\psfig{figure=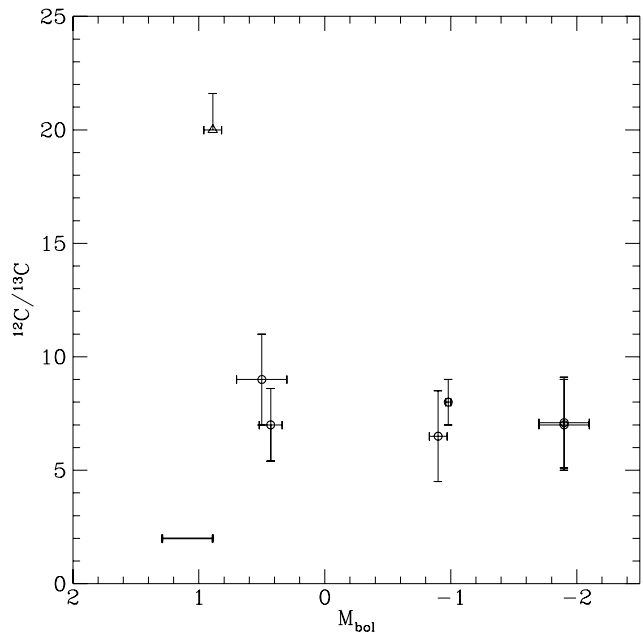,height=9cm}
}
\caption{The ratio of $\rm ^{12}C/^{13}C$ is plotted against the
bolometric magnitude of the program stars.  Uncertainties of $\pm$2 for the
isotope ratios are estimated for each star except for Arcturus, for which
$\pm$1 is estimated.
The absolute magnitudes are derived from the HIPPARCOS parallaxes.
Note that 47 Tuc 3501 and 4418 have both the same luminosity and
carbon isotopic ratio.
The horizontal line indicates the position of the bump in the
luminosity function of 47 Tuc (King \etal 1985).  }
\end{figure}

\begin{figure}
\centerline{
\psfig{figure=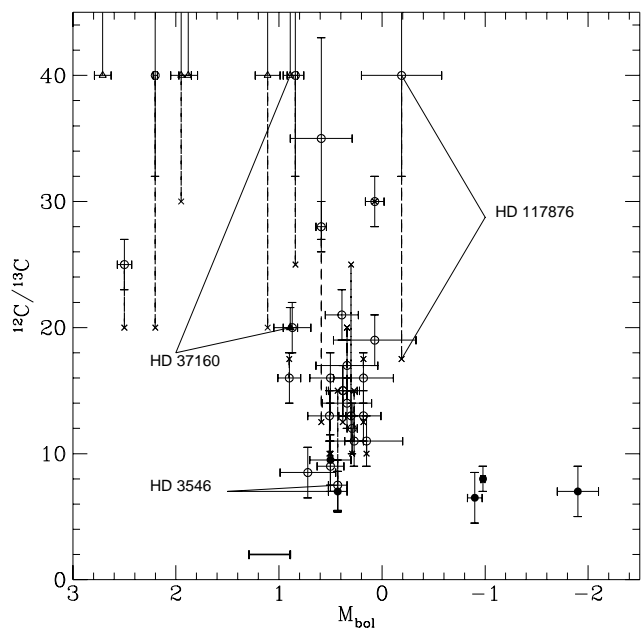,height=9cm}
}
\caption{Carbon isotopic ratios for the old disk giants by 
Cottrell \& Sneden (1986; crosses) and 
Shetrone et al.(1993; open circles and triangles)).
Our program stars are also shown (black circles and triangle).
Observations for the same stars are connected by broken lines.
The absolute magnitudes are derived from the HIPPARCOS parallaxes.
The horizontal line indicates the position of the bump in the
luminosity function of 47 Tuc (King \etal 1985).}
\end{figure}


\section {The onset of extra-mixing on the observed evolutionary
sequence}

Our observational results are summarized by referring to Figures 1 to 3.
We show in Figure 1 the observed dependence of $\rm ^{12}C/^{13}C$ ratio on
$\rm M_{bol}$.
This sequence for our sample of stars with moderate metal deficiencies
reveals a behavior of the surface $\rm ^{12}C/^{13}C$ ratio very similar to
the one previously observed at solar metallicity in M67.
Three main features appear:
(i) The least luminous star, $\phi^2$ Ori, shows a depletion of Li by
2.7 dex below the level accepted for young stars and a limit on the $\rm
^{12}C/^{13}C$ ratio in agreement with standard predictions for dilution
\footnote{(All the standard models computed by different authors give
very similar result for the post-dredge up value of 12C/13C; see for
example the compilation by Wasserburg et al. 1995. Boothroyd \& Sackman
(1997) showed that different initial carbon isotopic ratio do not affect 
significantly the post-dredge up result, whatever the metallicity of the 
models.)}
This indicates that the standard theoretical main sequence profiles of
$\rm ^{12}C$ and $\rm ^{13}C$ match the real chemical profiles, and that
the extra-mixing is only efficient after the completion of the
first dredge-up (see C94).
(ii) Then between $\rm M_{bol}$ = +0.9 and +0.5,
the observed isotopic ratio drops to
values near 7, well below the standard predicted post-dilution ratio.
It is interesting to note that
in the RGB luminosity function of 47 Tuc,
King, \etal (1985) localize the bump at
$\rm M_{bol} = +1.05 \pm 0.2$,
i.e. precisely in the region where the disagreement between
standard predictions and observations of the carbon isotopic ratio appears
in our sample.
This confirms that the extra-mixing which leads to very low
$\rm ^{12}C/^{13}C$ ratios in low-mass and metal-deficient evolved stars
becomes efficient exactly when the hydrogen-burning shell crosses the chemical
discontinuity created by the outward moving convective envelope.
(iii) Finally, from $\rm M_{bol}$ = +0.4 to $-$2,
there is no further change in the $\rm ^{12}C/^{13}C$ ratio.

The ``abrupt" change in  $\rm ^{12}C/^{13}C$ ratio shortly after
the luminosity-function bump occurs both at solar and at lower
metallicities,
although the final  $\rm ^{12}C/^{13}C$ ratio
is lower in the more metal-poor stars.
It is worth noting that the change in  $\rm ^{12}C/^{13}C$ ratio
is ``abrupt" only in terms of stellar luminosity.
The LFB is caused by a slower rate of evolution
for stars at this luminosity (that is, $dL/dt$ is smaller there)
owing to the H-burning shell contacting the H-rich, previously mixed, zone;
of the time stars take to evolve from the beginning of the LFB to the
tip of the giant branch, 15 --- 20\% of it
is spent in the LFB.
Consequently,
the change in the  $\rm ^{12}C/^{13}C$ ratio may not
be so abrupt in terms of time.

In order to enlarge our sample,
we recomputed the absolute magnitudes from the HIPPARCOS parallaxes
for the old disk giants with $\rm ^{12}C/^{13}C$ ratios derived by
SSP93.
These stars also have moderate metal deficiencies
(-1.0$\leq$[Fe/H]$\leq$-0.3) .
Due to the good luminosity determination, the complete data, which are
presented in Figure 2, can be reasonably viewed as an evolutionary sequence.
We also show the $\rm ^{12}C/^{13}C$ ratios previously derived by
Cottrell \& Sneden (1986) for the stars which belong to
the SSP93 sample.

The SSP93 isotope ratios tend to be systematically higher than the 
Cottrell \& Sneden
values for stars with high $\rm ^{12}C/^{13}C$,
but this is easily understood as being due to SSP93's 
superior spectral resolution.
The higher spectral resolution allows detection and better measurement
of the very weak $\rm ^{13}CN$ features.
This highlights a general problem in the determination of 
$\rm ^{12}C/^{13}C$ ratios: three effects operate to make $\rm ^{13}CN$
detection difficult for low-luminosity stars.
First, at lower luminosities the $\rm ^{13}CN$ proportion is lower
due to their unmixed state.  Second, model atmosphere effects operate
so that higher-gravity, lower-luminosity stars have weaker CN lines
even for constant chemical composition in the atmosphere.
Third, the lower stellar luminosity makes it more difficult to
obtain the requisite S/N to detect the weak lines.

\begin{figure}
\centerline{
\psfig{figure=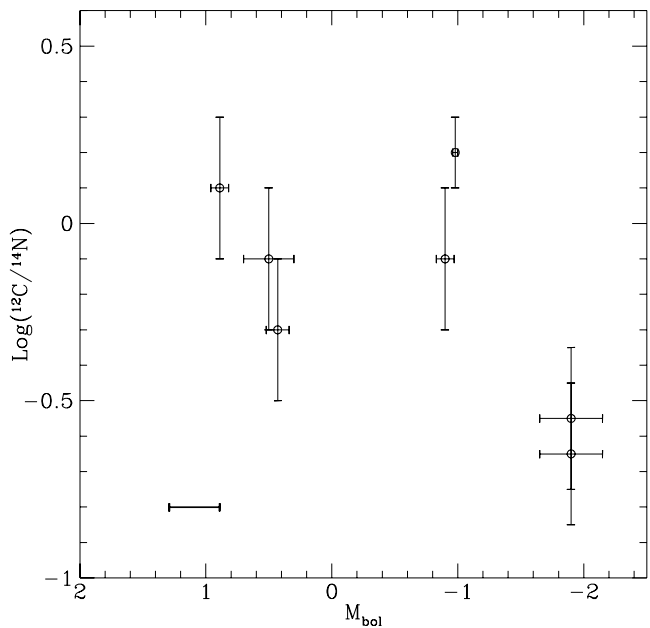,height=9cm}
}
\caption{Log $\rm ^{12}C/^{14}N$ against $\rm M_{bol}$ for the program
stars. 
The uncertainty in log $\rm ^{12}C/^{14}N$ is estimated to be 0.2 dex
except for Arcturus, for which it is probably 0.1 dex.
The horizontal line indicates the position of the bump in the
luminosity function of 47 Tuc (King \etal 1985).}
\end{figure}

Despite the uncertainties, it is clear that in the present complete
sample, no star with $\rm M_{bol} > +1.0$ 
shows a carbon isotopic ratio lower
than the one predicted by standard dilution models.
Figure 2 is strong evidence that extra mixing begins at a luminosity of
about $\rm M_{bol} = +1$.
This limit in luminosity corresponds to the approximative value for 
$\rm M_{bol}(bump)$ observed in clusters of the same [Fe/H] as the
present stars (Fusi Pecci \etal 1990).
The cutoff region appears broader here than in our restricted sample,
mainly due to a probable scatter in the stellar masses in SSP93 data.
Indeed, a 0.1M$_{\odot}$ difference corresponds approximatively to a shift
of 0.2 in $\rm M_{bol}$ for the position of the bump (Fusi Pecci \etal 1990).

We show the observed $\rm ^{12}C/^{14}N$ ratios plotted against
$\rm M_{bol}$ in Figure 3.
$\phi^2$ Ori presents a ratio consistent with standard post-dredge-up
predictions.
The $\rm ^{12}C/^{14}N$ ratio decreases simultaneously with the
$\rm ^{12}C/^{13}C$ ratio between $\rm M_{bol}$ = $+$1.0 and $+$0.5 by
about 0.3 dex, revealing that the extra-mixing process also slightly affects 
the N abundance.
Arcturus, for which the data is by far the best due to its brightness,
stands however above the curve for the other stars
(which really means above $\gamma$ Leo A, at the same luminosity, by 0.3
dex).
\footnote{Aoki and Tsuji (1997) have reanalysed the N abundance in 
Arcturus (and many cooler stars) and find log($^{12}$C/$^{14}$N)= 0.09,
which is within our uncertainty and somewhat closer to the 12C/14N 
ratio in Gamma Leo A.} 
This difference was recognized easily 15 years ago by
Lambert \& Ries (1981), who found a difference of 0.47 dex.
The 47 Tuc stars, at $\rm M_{bol} = -1.9$, present $\rm ^{12}C/^{14}N$
ratios lower than those in the less evolved objects.

In the M67 data (Brown 1987), a change in the C/N ratio at and above
the luminosity of the RGB bump is not clear.
This is probably due to errors, both random and systematic,
in the abundances derived from the synthetic spectrum fits,
and aggravated by the sparseness of the cluster RGB.
A systematic offset in the nitrogen abundance was found in that study,
and an attempt was made to calibrate this out in a post hoc way.
The result, however, is that any slow trend in C/N with luminosity
in the Brown (1987) data must be interpreted with caution.
It is more secure, however, to note that the total change in the
C/N ratio between stars at the base of the giant branch and those
at the RGB tip and in the post-helium core flash clump is larger
than predicted in standard models, so that the most luminous M67
giants again have also engaged in some extra mixing.

In our sample, the Li abundance drops between $\phi^2$ Ori
and $\varepsilon$ And (in terms of luminosity) after which it is below 
detectability.
It is difficult however to argue for the luminosity onset of extra-mixing 
from this data. Indeed, lithium destruction on the main sequence may lead 
to a dispersion of the abundances on the red giant branch which could
explain the difference between $\phi^2$ Ori and $\varepsilon$ And. 
In any case, the lithium abundances in the more evolved stars from our
sample clearly indicates that an extra-mixing mechanism transports lithium
from the convective envelope down to the region where it is destroyed by
proton capture after the end of the dilution phase.
In a large sample of evolved halo stars, Pilachowski, Sneden \& Booth
(1993) also showed that the lithium abundance continues to decrease
after the completion of the first dredge-up.
%
\section{Determination of a critical $\mu$-gradient from the
observed carbon isotopic ratios}

Recently, different groups have simulated extra-mixing between the
base of the convective envelope and the hydrogen burning shell in order to
reproduce the CNO abundances in RGB stars.
Denissenkov \& Weiss (1995) modeled this deep mixing by adjusting both the
mixing depth and rate in their diffusion procedure.
Wasserburg \etal (1995) and Boothroyd \& Sackman (1997)
used an ad-hoc ``conveyor-belt" circulation model, where
the depth of the extra-mixed region is related to a parametrized temperature
difference up to the bottom of the hydrogen-burning shell.
On the other hand, other authors attempted to relate the extra-mixing with
physical processes, among which rotation seems to be the most promising.
Sweigart \& Mengel (1979) suggested that meridional circulation on
the RGB could lead to the low $^{12}$C/$^{13}$C observed in field giants.
Taking into account the interaction between meridional circulation
and turbulence induced by rotation in stars, Charbonnel (1995)
showed that rotation-induced mixing can account for the observed
behavior of carbon isotopic ratios and for the Li abundances in
Population II low mass giants.

In these different approaches, the common underlying question is the
determination of the extension of the region where extra-mixing occurs,
and more precisely the nature of the stabilizing factor.
An investigation into the mechanism of extra mixing
must include an understanding of why the mechanism does {\it not}
operate at luminosities below that of the RGB bump, and what is the
nature of the factor that stabilizes the star against extra mixing
before it reaches this point.
When suggesting that the extra-mixing on the RGB could be related to
rotation, Sweigart \& Mengel (1979) and Charbonnel (1994, 1995)
underlined the inhibiting effect of molecular weight (or $\mu$) barriers.
Indeed, as argued by Mestel (1957), gradients of molecular weight tend to
restrain the circulation and to stabilize the mixing processes related to
rotation.
We will show now how 
the observations gathered in this paper allow us to constrain, from
simple considerations, the conditions in which a $\mu$-barrier shields the
central regions of a star from extra-mixing.

We have computed some standard stellar models for exploration of these
questions.
The evolutionary sequences were followed from the pre-main sequence
up to the RGB tip with the Toulouse-Geneva code
(see Charbonnel \etal 1992).
The input microphysics is the same as used in Charbonnel et al. 
(1997). The relative ratios for the heavy elements correspond to the
mixture by Grevesse \& Noels (1993). We take the same isotopic ratios as
Maeder (1983). 
Neither element segregation or rotation-induced mixing are included in the
present computations.

\begin{figure}
\centerline{
\psfig{figure=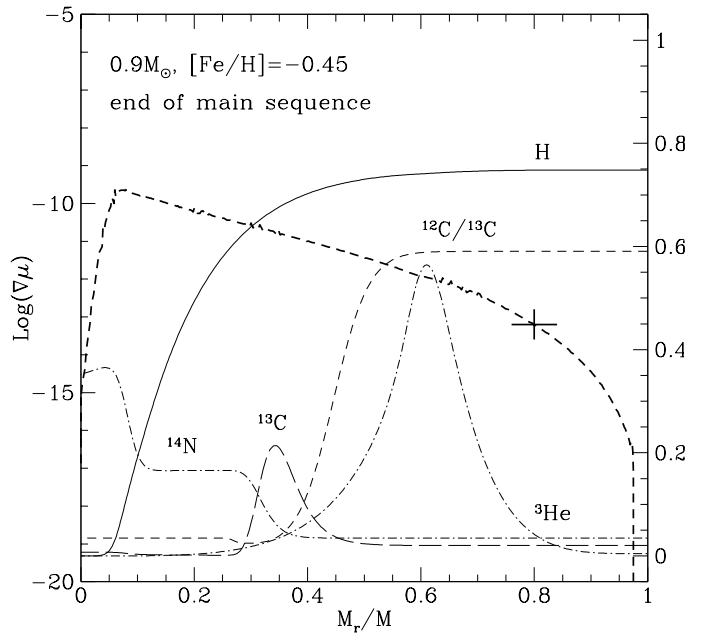,height=9cm}
}
\caption{
Molecular weight gradient (in log scale, bold dashed line, left axis)
and composition profiles (right axis)
in a 0.9M$_{\odot}$, [Fe/H]=-0.45 stellar model,
at the end of the main sequence.
The mass fractions are multiplied by 100 for $^3$He and $^{14}$N, by
1000 for $^{13}$C; the ratio $^{12}$C/$^{13}$C is divided by 100.
The thick cross corresponds to $(\nabla ln \mu)_c$,
the critical molecular weight gradient
obtained with the prescription by Huppert \& Spiegel (1977) discussed in
\S 5}
\end{figure}

\begin{figure}
\centerline{
\psfig{figure=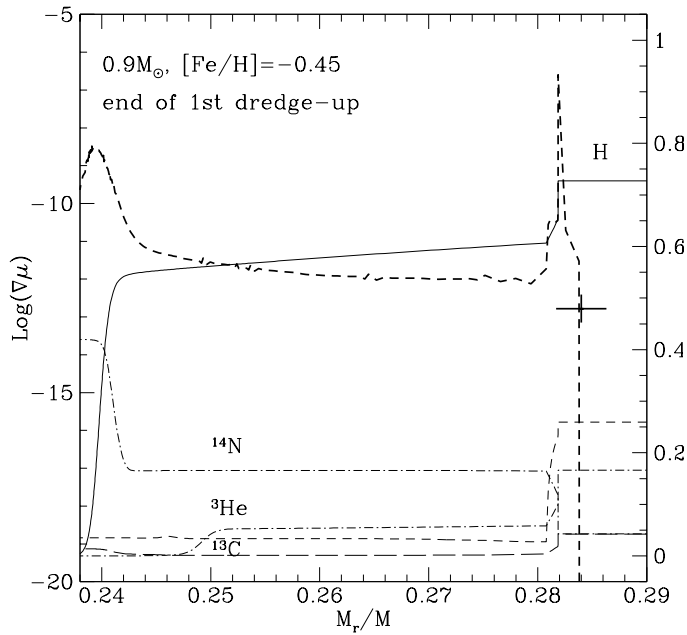,height=9cm}
}
\caption{Same as Fig. 4, after the completion of the first dredge-up,
in the H-burning shell region.
The base of the convection zone is located at M$_r$/M$\simeq$0.283.}
\end{figure}

\begin{figure}
\centerline{
\psfig{figure=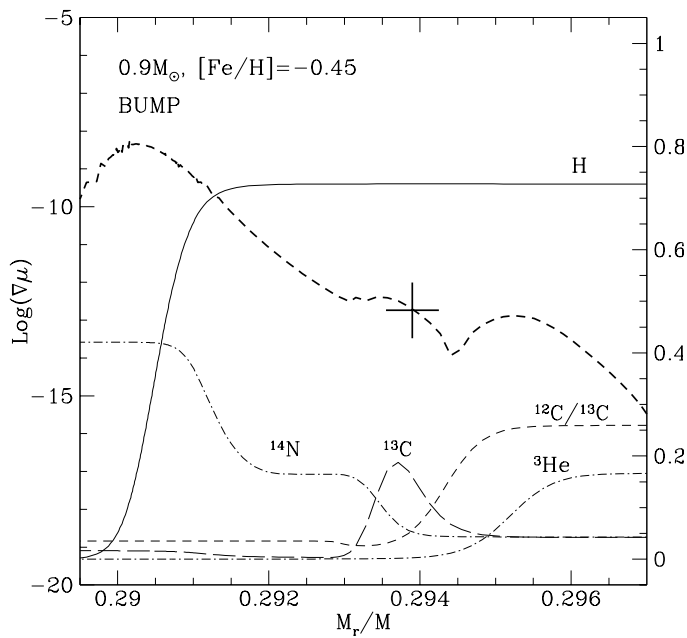,height=9cm}
}
\caption{Same as Fig. 4, after the RGB bump, in the H-burning shell region.
The base of the convection zone is located at M$_r$/M$\simeq$0.313.}
\end{figure}

The gradients of molecular weight ($\nabla \ln \mu = d \ln \mu /dr$)
in a standard stellar model of 0.9M$_{\odot}$ with $\rm [Fe/H] = -0.45$,
typical of the present observed sample, are presented
in Figures 4, 5 and 6 at three different evolutionary stages:
at the end of the main sequence, after the completion
of the first dredge-up, and shortly after the RGB bump.
We also show the abundance profiles for H, $^3$He, $^{13}$C and $^{14}$N,
and the $^{12}$C/$^{13}$C ratio.

While on the main sequence, nuclear burning leads to an increase of the
mean molecular weight in the central regions of the star (see Figure 4).
During the first dredge-up phase, the convective envelope homogenizes
the star down to very deep regions, and builds a very steep gradient
of molecular weight at the point of its maximum penetration
($\nabla \ln \mu \simeq 10^{-6}$ at ${\rm M}_r/{\rm M} \simeq 0.282$
in Figure 5).
Further on, as the star keeps ascending the RGB, the hydrogen burning
shell becomes thinner and moves outwards in the mass scale, while
the convective envelope retreats.
As discussed in \S 3, our observations provide strong evidence that
the extra-mixing is inhibited until the star reaches the LFB.
The shape of $\nabla \ln \mu$ just after this evolutionary point,
i.e. after the encounter of the hydrogen burning shell with the previously
mixed region, is shown in Figure 6.
Below the base of the convective envelope
$\nabla \ln \mu$ is much lower at this time.

Our observations provide a precise clue on the extension of the
extra-mixed region down to the nuclear burning layers.
The low $^{12}$C/$^{13}$C ratios we see in our sample ($\simeq$ 7)
can indeed be reached if the extra-mixing develops down to 
M$_r$/M$\simeq$0.294 in our 0.9M$_{\odot}$ model (see Figure 6).
At this point, $\nabla \ln \mu \simeq 1.5 \times 10^{-13}$.
We now assign this value to ($\nabla \ln \mu)_{c,obs}$,
that we define as the ``observational" critical $\mu$-gradient
which appears to shield the central regions of the star from extra-mixing.
In our 0.9M$_{\odot}$ model, the temperature difference $\Delta \log {\rm T}$ 
between the bottom of the hydrogen burning shell and this point
is equal to 0.26.
This value for $\Delta \log {\rm T}$ is in agreement with the one
Boothroyd \& Sackman (1997) have to impose in their parametrized mixing model
in order to match the observed $^{12}$C/$^{13}$C ratios at solar metallicity.

Let us check now to what region
($\nabla \ln \mu)_{c,obs} \simeq 1.5 \times 10^{-13}$ corresponds, 
in terms of expected $^{12}$C/$^{13}$C, at different evolutionary stages, 
and for stars of different metallicities.
We see in Figure 4 that while the star is on the main sequence,
the real $\nabla \ln \mu$ becomes higher than
($\nabla \ln \mu)_{c,obs}$ in the very external part of the star,
around M$_r$/M equal to 0.75.
If we assume that such a $\mu$-barrier can not be penetrated, then
on the main sequence no extra-mixing is expected to occur in the
stellar region of energy production.
This explains the observations of the carbon isotopic ratio
in $\phi^2$ Ori, in the less luminous stars of SSP93's sample, 
and in the less evolved RGB stars of M67 (GB91), where
we see a perfect agreement with standard theoretical dilution.
This is also in agreement with helioseismological constraints.
Indeed, solar structure computations bring important information on the
critical $\mu$-gradients that limit the mild mixing which is necessary to
explain the lithium depletion in the Sun.
Richard \etal (1996) and Richard \& Vauclair (1997) showed that the best solar
models (for what concerns helioseismological comparison and agreement with Li
and Be observations) are those including both element segregation and
rotation-induced mixing, where the mixing is cut-off when the
$\mu$-gradient becomes $\geq$ to 1.5 - 4 $\times 10^{-13}$.
It is highly satisfactory to get the same value for ($\nabla \ln
\mu)_{c,obs}$ \
from two completely different observational constraints.

\begin{figure}
\centerline{
\psfig{figure=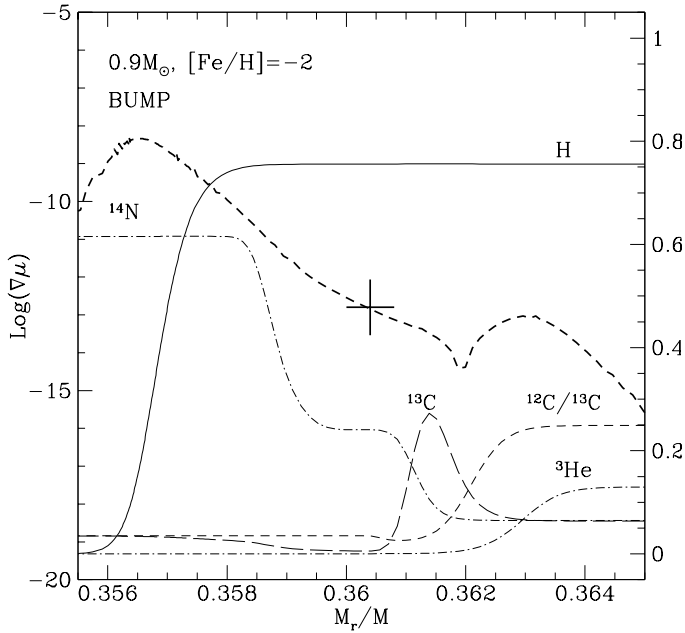,height=9cm}
}
\caption{Same as Fig. 6 in the 0.9M$_{\odot}$, [Fe/H]=-2 stellar model.
The base of the convection zone is located at M$_r$/M$\simeq$0.387.
The mass fractions are multiplied by 100 for $^3$He, and by 50000 and
5000 respectively for $^{13}$C and $^{14}$N;
the ratio $^{12}$C/$^{13}$C is divided by 100}
\end{figure}

\begin{figure}
\centerline{
\psfig{figure=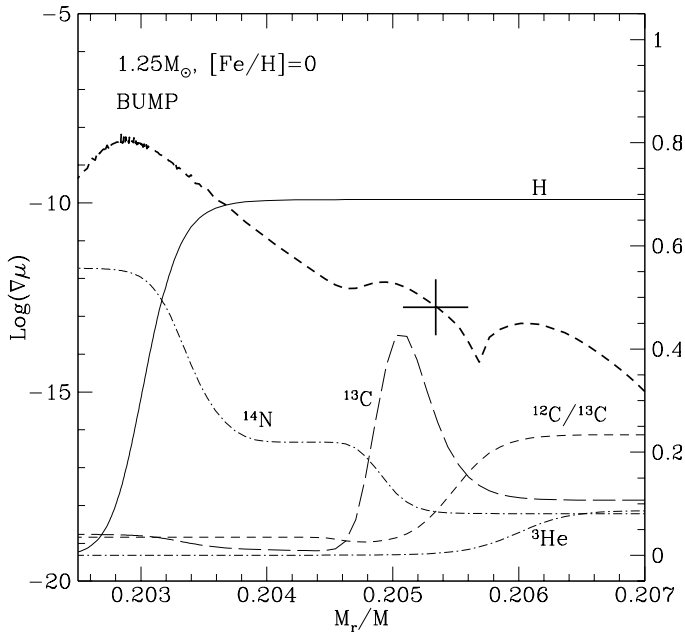,height=9cm}
}
\caption{Same as Fig. 6 in the 1.25M$_{\odot}$, [Fe/H]=0 stellar model.
The base of the convection zone is located at M$_r$/M$\simeq$0.218.
The mass fractions are multiplied by 100, 1000 and 50 respectively
for $^3$He, $^{13}$C and $^{14}$N;
the ratio $^{12}$C/$^{13}$C is divided by 100}
\end{figure}

The gradients of molecular weight shortly after the RGB bump
in a model of 0.9M${\odot}$ with [Fe/H] = $-2$ (typical of a globular
cluster giant)
and in a model of 1.25M${\odot}$ at solar metallicity
(typical of a M67 giant star)
are presented respectively in Figures 7 and 8.
If we assume that the extra-mixing reaches down to the same
$(\nabla \ln \mu)_{c,obs}$ whatever the star,
then the $^{12}$C/$^{13}$C is expected to decrease down to its equilibrium
value for the globular cluster giant, and down to $\simeq$ 12 in the 
open cluster giant.
These values, and the dependency with metallicity, are exactly
as observed in the giants in open clusters by Gilroy (1989)
and are in agreement with the other observations of field stars
and globular cluster giants.
They are independent of any modeling for the extra-mixing.
It appears then that the value we derived empirically for 
$(\nabla \ln \mu)_{c,obs}$ is ``universal", in the sense that an unique
value is sufficient to describe different constraints.

As already shown by C94, if molecular weight barriers are the inhibiting
factor, then no extra-mixing is expected to occur in stars which ignite
helium in a non-degenerate core, i.e. in stars with masses higher
than 1.7 to 2.2M$_{\odot}$ (the exact value depends on the metallicity).
Indeed, in these more massive stars, the hydrogen burning shell does not
have the time to reach the regions that have been homogeneized during the
first dredge-up, so that $(\nabla \ln \mu)$ below the convective
envelope is always higher than $(\nabla \ln \mu)_{c,obs}$.
Under these conditions, extra-mixing is inhibited in these stars.
This is confirmed by the observational data in open clusters with 
turn-off masses higher than about 2M$_{\odot}$ (Gilroy 1989).

\section{Clues on the nature of the extra-mixing process on the
rgb}

The results described previously were obtained using observational constraints
only, and did not rely on any special prescription for the extra-mixing.
We can nevertheless check whether the value we get for
$(\nabla \ln \mu)_{c,obs}$ is consistent with what is expected in a very
simplified rotation framework.
Let us follow the suggestion by
Huppert \& Spiegel (1977) according to which meridional currents can
penetrate into a region with a stable gradient of molecular weight
within a scale height given by
$h \simeq r {{\Omega(r)}\over{N_{\mu}}}$,
where r is the local radius,
$\Omega(r)$ is the angular rotation velocity,
and N$_{\mu}$ the buoyancy frequency due to the $\mu$-gradients.
We can derive then a critical $\mu$-gradient by specifying that h must be a
small fraction of r ($\rm h = \epsilon r$):
$(\nabla ln \mu)_{c,theor} \simeq {1\over{\epsilon ^2}}
{{r^2 \Omega^2(R_c)}\over {G M(r_c)}}$,
where all the quantities are computed at the place where the actual
$\nabla \ln \mu$ is equal to the critical one.

For each of our 3 models, we computed $(\nabla \ln \mu)_{c,theor}$
for a stellar rotation velocity of 2.5 km sec$^{-1}$,
typical on the RGB for the stellar masses in consideration 
(De Medeiros \etal 1996).
The $(\nabla \ln \mu)_{c,theor}$ we obtain in this model-dependent approach
are shown in Figures 4 to 8 (thick cross).
They are in perfect agreement with the constraints coming from the
observations discussed in \S 4.
This result tends to indicate that the extra-mixing process on the RGB
is related to rotation.

However, we have to be cautious in our conclusions.
Indeed, $(\nabla ln \mu)_{c,theor}$ is derived here in a simplified
framework.
Complete computations with rotation-induced mixing have to be performed. 
The transport of chemicals and angular momentum should be treated 
simultaneously to properly handle the development of $\mu$-barriers 
(despite rotational mixing) during stellar evolution.  
Different stellar rotational histories also have to be considered, which
may result in different efficiencies of the extra-mixing, and lead
to a dispersion of chemical anomalies from star to star.
Observational support for this approach may exist from the variations of 
oxygen abundances and rotational velocities in blue horizontal branch 
stars in three globular clusters (Peterson \etal 1995).

\section{Conclusions}

We have assembled accurate $\rm ^{12}C/^{13}C$ ratios, as well as 
abundance data for C, N, O and Li for five field giants with 
[Fe/H] $\simeq - 0.5$ including Arcturus, and for two stars in the 
globular cluster 47 Tuc, and reviewed previously published data for a 
number of other field giants.
Using HIPPARCOS parallaxes to derive $\rm M_{bol}$ values, we placed these
stars into an evolutionary sequence along the ascent of the RGB.
The sequence shows that the $\rm ^{12}C/^{13}C$ ratios drops from 
$\sim 20$ to near 7 between $\rm M_{bol}$ = +0.9 and +0.5, while Li 
disappears.
The C/N ratio is also affected, with log ($\rm ^{12}C/^{14}N$) 
diminishing from +0.1 to $-$0.6 between $\rm M_{bol} =$ +0.9 and $-$2.
Arcturus stands above the curve defined by the other stars by 0.4 dex.
These observations confirm the existence of an extra-mixing process that
becomes efficient on the RGB only when low-mass stars reach the so-called
luminosity-function bump. This result is established for the first time
for stars with moderate metal deficiencies. 

We used the values of the carbon isotopic ratio observed in our sample
to get constraints on the $\mu$-barriers that may shield the central regions
of a star from extra-mixing.
We showed that the same value of the critical gradient of molecular weight
leads to $\rm ^{12}C/^{13}C$ ratios observed at different metallicities.
This observational critical $\mu$-gradient is in very good agreement with the
critical $\mu$-gradient which is expected to stabilize meridional circulation.
This result brings strong clues on the nature of the extra-mixing which
occurs on the RGB, and indicates that it is related to rotation.
Observational support for this approach may exist from the variations
of oxygen, sodium and aluminum in globular cluster red giants as well
as the measurement of rotational velcoities in blue horizontal branch 
stars in three globular clusters (Kraft, 1994; Kraft et al., 1997; 
Peterson et al. 1995).

We thank the Institute for Nuclear Theory at the University of
Washington where this work was initiated.
C.C. is most appreciative for the hospitality shown to her while a
visitor at the Space Telescope Science Institute.
G.W. and J.A.B. would like to acknowledge the support of the Kennilworth
Fund of the New York Community Trust.
Some of the spectra were obtained by Guillermo Gonzalez.
The spectra analyzed as part of this project were obtained
at the Dominion Astrophysical Observatory, 
Herzberg Institute of Astrophysics,
National Research Council, Canada.

{}

\end{document}